\documentclass[useAMS,usenatbib]{mn2e}
\usepackage{url,times,graphicx,amsmath,amsfonts,amssymb,aas_macros,color,epsfig,epstopdf}
\newcommand{\nil}[1]{#1}

\newcommand{\Tab}[1]{Table~\ref{#1}}

\newcommand{\Fig}[1]{Fig.\ref{#1}}

\newcommand{\hMpc}{{\ifmmode{h^{-1}{\rm Mpc}}\else{$h^{-1}$Mpc}\fi}}
\newcommand{\hkpc}{{\ifmmode{h^{-1}{\rm kpc}}\else{$h^{-1}$kpc}\fi}}
\newcommand{\hMsun}{{\ifmmode{h^{-1}{\rm {M_{\odot}}}}\else{$h^{-1}{\rm{M_{\odot}}}$}\fi}}

\newcommand{\ltsima}{$\; \buildrel < \over \sim \;$}
\newcommand{\gtsima}{$\; \buildrel > \over \sim \;$}
\newcommand{\lsim}{\lower.5ex\hbox{\ltsima}}
\newcommand{\gsim}{\lower.5ex\hbox{\gtsima}}

\def\lesssim{\mathrel{\hbox{\rlap{\hbox{\lower4pt\hbox{$\sim$}}}\hbox{$<$}}}}
\def\gtrsim{\mathrel{\hbox{\rlap{\hbox{\lower4pt\hbox{$\sim$}}}\hbox{$>$}}}}

\title[Renegade Subhaloes in the Local Group]
      {Renegade Subhaloes in the Local Group}
\author[Knebe et al.] 
{Alexander Knebe$^1$, Noam I Libeskind$^2$, Timur Doumler$^{2,3}$, Gustavo Yepes$^1$, \newauthor Stefan Gottl\"ober$^2$, Yehuda Hoffman$^4$
  \\
  $^1$Grupo de Astrof\'\i sica, Departamento de Fisica Teorica, Modulo C-15, Universidad Aut\'onoma de Madrid, Cantoblanco E-28049, Spain\\
  $^2$\nil{Leibniz Institut f\"{u}r Astrophysik,} An der Sternwarte 16, D-14482 Potsdam, Germany\\
 $^3$Universite Lyon 1, CNRS/IN2P3/INSU, Institut de Physique Nucleaire, 69622 Villeurbanne, Lyon, France\\
 $^4$Racah Institute of Physics, The Hebrew University of Jerusalem,  Jerusalem 91904, Israel
  }

\begin{document}


\pagerange{\pageref{firstpage}--\pageref{lastpage}} \pubyear{2008}

\maketitle

\label{firstpage}

\begin{abstract}
  Using a dark matter only Constrained Local UniversE Simulation (CLUES) we examine the existence of subhaloes that change their affiliation from one of the two prominent hosts in the Local Group (i.e. the Milky Way and the Andromeda galaxy) to the other, and call these objects "renegade subhaloes". In light of recent claims that the two Magellanic Clouds (MCs) may have originated from another region (or even the outskirts) of the Local Group or that they have been spawned by a major merger in the past of the Andromeda galaxy, we investigate the nature of such events. However, we cannot confirm that renegade subhaloes enter as deep into the potential well of their present host nor that they share the most simplest properties with the MCs, namely mass and relative velocity. Our simulation rather suggests that these renegade subhaloes appear to be flying past one host before being pulled into the other. A merger is not required to trigger such an event, it is rather the distinct environment of our simulated Local Group facilitating such behavior. Since just a small fraction of the full $z=0$ subhalo population are renegades, our study indicates that it will be intrinsically difficult to distinguish them despite clear differences in their velocity, radial distribution, shape and spin parameter distributions.
  
\end{abstract}


\section{Introduction}
\label{sec:introduction}
In the concordance cosmology, structure forms in a hierarchical, ``bottom-up'' fashion that leads to the accretion of small substructures by large dark matter haloes. The current paradigm holds that substructures orbit within their parent haloes until tidal stripping rips them apart as they sink to the centre of the potential by dynamical friction.

The Local Group of galaxies is an ideal testbed for this kind of near-field cosmology \citep{Freeman02} as the observational data is better than for any distant galaxy. But as already prompted before, how certain can we be that the Milky Way (MW) or the Andromeda galaxy (M31) are in fact typical galaxies of their mass or luminosity \citep[cf. ][]{Forero-Romero11}?  These two  galaxies are the dominant members of the Local Group and form a two-body system on a collision course in approximately 2-3 Gyrs \citep{Cox08,Hoffman07}. Their proximity at the present time indicates that they may already have started to influence each other or their respective satellite populations. Perhaps this influence is related to the (in)famous ``disk of satellites'' \citep[e.g.][]{Metz08}; there is also the question of how frequently the largest companions  of the MW - the Small and the Large Magellanic Clouds (SMC and LMC) - are found. A recent study by \citet{Liu10} indicates that only of order 3\% of Milky Way type galaxies in the SDSS Data Release 7 host two satellites with luminosities similar to the Magellanic Clouds (MC). This result is also supported by \citet{James10} who investigated 143 luminous spirals in H$\alpha$ finding that the MW is an unusual galaxy both for the luminosity and the proximity of its two brightest satellites. Therefore, both these satellites are often considered outliers with respect to the full system of satellites orbiting the MW and their formation scenario is a matter of debate \citep{Besla07, Peebles09, Kallivayalil09, Metz09}.

There are several investigations that suggest that the LMC and SMC are on their first infall as suggested by recent proper motions measurements \citep{Besla07, Tollerud11} -- as opposed the to recent claims by \citet{Sales11} that the LMC is not necessarily on its first approach to the MW based upon the Aquarius simulation \citep{Springel08}; if they are only loosely (if at all) bound satellites of the MW then where did they come from? \citet{Yang10} deal with the possibility that both the Magellanic Clouds have been expelled from M31 due to a major merger event. The scenario envisioned by them is that they are ejected tidal dwarf galaxies from a previous major merger occurring at the M31 location \citep{Hammer10}. But there is also the notion that they may simply be accreted objects that formed in the outer reaches of the Local Group \citep{vandenBergh10, Busha10}. While these studies are still speculative, they nevertheless show that M31 may play a vital role in shaping the orbits of the Magellanic Clouds \citep[see e.g.][]{Kallivayalil09}. Neither the MW nor M31 may be understood in isolation but only in the context of the (unique) environment of the Local Group and its formation.

In this \textit{Letter} we use a Constrained Local UniversE Simulation (\textit{http://www.clues-project.org}) to search for events where subhaloes change their host halo affiliation. These simulations are well suited for this idea as they directly model the Local Group (consisting of the two-body system MW and M31) within the correct environment and in a cosmological framework of the concordance model. We investigate the likelihood that subhaloes were under the influence of one of the two hosts at some previous time while at $z=0$ are within the virial radius of the other host. We colloquially call them ``renegade'' or ``disloyal''  subhaloes. 

\section{The Simulations}
\label{sec:simulations}

\paragraph*{Constrained Simulation of the Local Group}

We choose to run a dark matter only simulation using standard $\Lambda$CDM initial
conditions, that assume a WMAP5 cosmology \citep{Komatsu09}, i.e.
$\Omega_m = 0.233$, $\Omega_{b} = 0.046$, $\Omega_{\Lambda} = 0.721$. We
use a normalization of $\sigma_8 = 0.817$ and a $n=0.96$ slope of the
power spectrum. We use the treePM-SPH code \texttt{GADGET2}
\citep{Springel05} to simulate the evolution of a cosmological box
with side length of $L_{\rm box}=64 h^{-1} \rm Mpc$. Within this box
we identified (in a lower-resolution run utilizing $1024^3$ particles)
the position of a model local group that closely resembles the real
Local Group \citep[cf.][]{Libeskind10}. This Local Group has then been
re-sampled with 64 times higher mass resolution in a region of $2
h^{-1} \rm Mpc$ about its centre giving a nominal resolution
equivalent to $4096^3$ particles giving a mass resolution of $m_{\rm
  DM}=2.1\times 10^{5}$\hMsun. For more details
we refer to the reader to \citet{Gottloeber10}. 

\paragraph*{The (Sub-)Halo Finding and Tracking}
We use the MPI+OpenMP hybrid halo finder \texttt{AHF} \citep[][]{Knollmann09} to identify haloes and subhaloes in our simulation which is the successor of the \texttt{MHF} halo finder by \citet{Gill04a}. We only consider subhaloes where the number of gravitationally bound particles is larger than 20.

We build merger trees by cross-correlating haloes in consecutive simulation outputs. The direct progenitor at the previous redshift is the object that shares the most particles with the present halo \textit{and} is closest to it in mass. For more details we point to the reader to e.g. \citet{Libeskind10} or \citet{Knebe10a}.

\paragraph*{Seeking Renegade Subhaloes}
The prime focus of this \textit{Letter} is to examine the set of subhaloes that change their host affiliation. These subhaloes are identified as being within a fixed distance of one host at some early time and within the same fixed distance of the other host at $z=0$. In practice we identify renegade subhaloes within $D = R_{\rm vir}^{\rm host}$ of the two main members of the Local Group (MW and M31) at redshift $z=0$. Each subhalo is then traced back in time using the merger tree. At each snapshot we compare its distance to the other host and note when this distance falls below $D(z) = R_{\rm vir}^{\rm host}(z)$, i.e. when it is accreted, recording it as a ``host change''.

\section{Results}
\label{sec:results}
\begin{figure}
\noindent
\centerline{\hbox{\psfig{figure=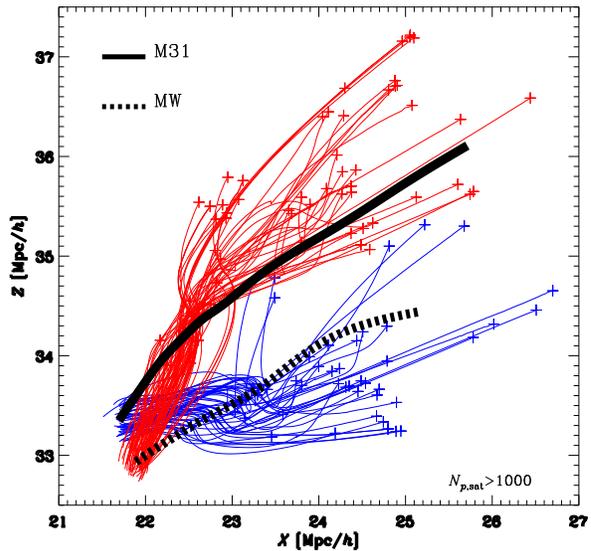,width=1.\hsize,height=0.92\hsize,angle=0}}}
\caption{Projection of the flights paths in comoving
  coordinates of both hosts MW (thick dashed) and M31 (thick solid) as
  well as those satellites that changed loyalty. The crosses mark the subhaloes' staring points at high redshift whereas the end points of $z=0$ are unmarked. Please note that we only show the trajectories for subhaloes with more than 1000 particles.
  }
\label{fig:orbit}
\end{figure}

We begin this section by describing how disloyalty in subhaloes may arise. In  \Fig{fig:orbit} we show renegade subhaloes of our simulated local group. The two thick lines represent the flight paths of the "MW" (dashed) and "M31" (solid) from $z=5$ until today, from now on simply referred to as MW and M31: they are currently approaching each other with 113 km/s, having a (relative) transversal velocity of 23 km/s. In blue and red we show the trajectories of renegade subhaloes of the simulated MW and M31, respectively. Please note that this figures serves merely an illustrative purpose and hence has been limited to just the most massive subhaloes, those with more than 1000 particles. Including lower mass subhaloes has no qualitative effect. We actually observe the same phenomenon reported and quantified for the CLUES simulation within a WMAP3 framework: (renegade) subhaloes fall in to their respective (first) hosts from preferred directions, i.e. they are residing inside the filamentary structure through which the two hosts travel towards the (simulated) Virgo cluster \citep{Libeskind10infall}. However, while this filament is broken into two strands in the WMAP3 simulation, a visual impression (not presented here) of the WMAP5 simulation reveals that the MW and M31 are occupying the same filament (Libeskind et al., in preparation). This explains the possibility to exchange subhaloes whereas we cannot confirm their existence in the WMAP3 simulation (see below).

We find that the situation is slightly asymmetric with respect to the two hosts: there are 128 renegades in total: 72 of the MW and just 57 of M31 when considering only well resolved subhaloes (i.e. those with more than 20 particles) found by our halo finder \texttt{AHF}. Note that we fail to find multiple host changes: if a subhalo changes its host it happens just once (or never) in our particular simulation. The number of renegade subhaloes (and their fractions with respects to the total number of subhaloes) is summarized in \Tab{tab:numbers}.

We apply our algorithm at higher redshift when the two haloes were further apart; note that the two hosts monotonically approach each other since $z\approx5$. We find that the number of renegade subhaloes is a function of host distance - the closer the two haloes are to each other, the more renegades exist. We thus conclude that the number of renegades is an indicator of the continuous approach of the MW and M31 and foretells their imminent merger, respectively \citep{Hoffman07}. We repeated our analysis in the WMAP3 CLUES simulation \citep[e.g.][]{Libeskind10,Knebe11a} where the $z=0$ separation of the two hosts is~$\sim$ 700\hkpc. In this case we find no renegades. We ascribe the decreased number to the lower $\sigma_8$-normalisation of that particular model and the smaller masses of the two respective hosts, respectively. Further, whereas the two hosts appear to be flying within one common filament in the WMAP5 simulation only gradually moving towards each other over time, the orbital approach in the WMAP3 simulation is more rapid close to redshift $z=0$ due to the embedding of each host in a separate filament (cf. Libeskind et al., in preparation).

\begin{table}
 \caption{The number of renegade satellites (with $N_{p,\rm sat}\geq20$) as a function of redshift $z$ and distance $D_{\rm MW-M31}$ (as measured in \hkpc), respectively. The last line summarized the percentage $p$ of renegade subhaloes with respects to the total number of subhaloes.}
\begin{center}
\begin{tabular}{lcllllll}
\hline
$z$                & =  & 0.0 & 0.05 & 0.16 & 0.23 & 0.29 & 0.68\\
$D_{\rm MW-M31}$   & =  & 545 & 600  & 700  & 750  & 800  & 1000\\
$N$                & =  & 128 & 107  &  26  &   7  &   1  &    0\\
$p$                & =  & 5\% & 4\%  & 1\%  &$<$1\%&$<$1\%& 0\%\\
\hline
\end{tabular}
\end{center}
\label{tab:numbers}
\end{table}

\begin{figure}
\noindent
\centerline{\hbox{\psfig{figure=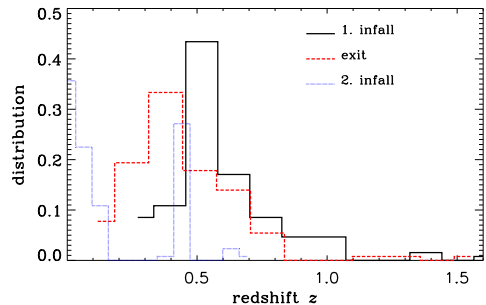,width=0.85\hsize,angle=0}}}
\caption{The distribution of first infall, exit and second infall redshifts for all renegade subhaloes.}
\label{fig:Pzred}
\end{figure}

While \Fig{fig:orbit} shows the trajectories of the renegade subhaloes it does not provide information about the first (host) infall, exit and second (host) infall times. To gain insight into this issue we present in \Fig{fig:Pzred} the corresponding redshift distributions for the combined sample of the MW and M31: a peaked distribution can be indicative of group infall as reported before by \citet{Li08subgroups,Angulo09,Klimentowski10}. Here we find that even the combined sample shows a peaked distribution meaning that infall and exit happened for both populations of renegades (i.e. MW and M31 renegades) at approximately the same time. We find that (most of) the renegades fell in at approximately $z\approx0.55$, left it shortly afterwards ($z\approx0.35$) to enter the other host close to redshift $z=0$. This confirms the picture already drawn from \Fig{fig:orbit}: renegade subhaloes come from the same region \citep[cf.][]{Libeskind10infall}, fly past one host, and are (gravitationally) pulled over to the other host, ending up within the area of trade of the latter at redshift $z=0$. It is important to note that despite claims by, for example, \citet{Fouquet11} that a major merger is required to trigger renegade satellites, we find them without such events. In our simulation the constantly decreasing distance and proximity, respectively, between the two hosts and the existence of backsplash galaxies \citep[i.e. galaxies that enter and leave the virial radius of a host halo; cf.][]{Gill05,Knebe11a} is sufficient to explain the appearance of renegade subhaloes.

\begin{figure}
\noindent
\centerline{\hbox{\psfig{figure=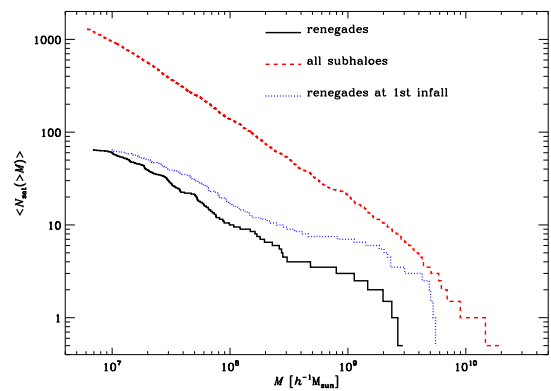,width=0.9\hsize,angle=0}}}
\caption{The mass function of renegade subhaloes at redshift $z=0$ (solid line) and at their respective infall time into the first host (dotted line) combined for M31 and the MW. We also show the (combined) mass function for all present-day subhaloes (dashed line).}
\label{fig:massfunc}
\end{figure}

The next question we address is that of mass distribution of the renegade subhaloes. The corresponding plot is to be found in \Fig{fig:massfunc} where we show the mass spectrum of the disloyal subhaloes at redshift $z=0$ (solid line) in comparison to the full subhalo mass function (dashed line), as an average of the MW and M31. We additionally show the distribution of renegade halo masses at the time of the initial infall into their first host as dotted line.

We observe that both mass functions at redshift $z=0$ follows a power-law. Or in other words, the renegade function is not peaked at a particular mass. However, the masses of the renegades are systematically lower than those of their loyal companions. Further, it is also apparent the renegades are less massive than the LMC yet compatible with the SMC. Our simulations therefore provide evidence for the existence of renegades when the environmental setting of the hosts is just about right, but the link to the Magellanic Clouds is not immediately apparant. That being said, our inability to produce MC-like renegade subhaloes may just be due the difficulty in forming such massive satellites in the first place \citep[e.g.][]{Boylan-Kolchin10,Boylan-Kolchin11,Sales11}. \Fig{fig:massfunc} further confirms that the renegades suffered mass loss as their original masses at first infall were around a factor of two larger.

\begin{figure*}
\noindent
\centerline{\hbox{
\psfig{figure=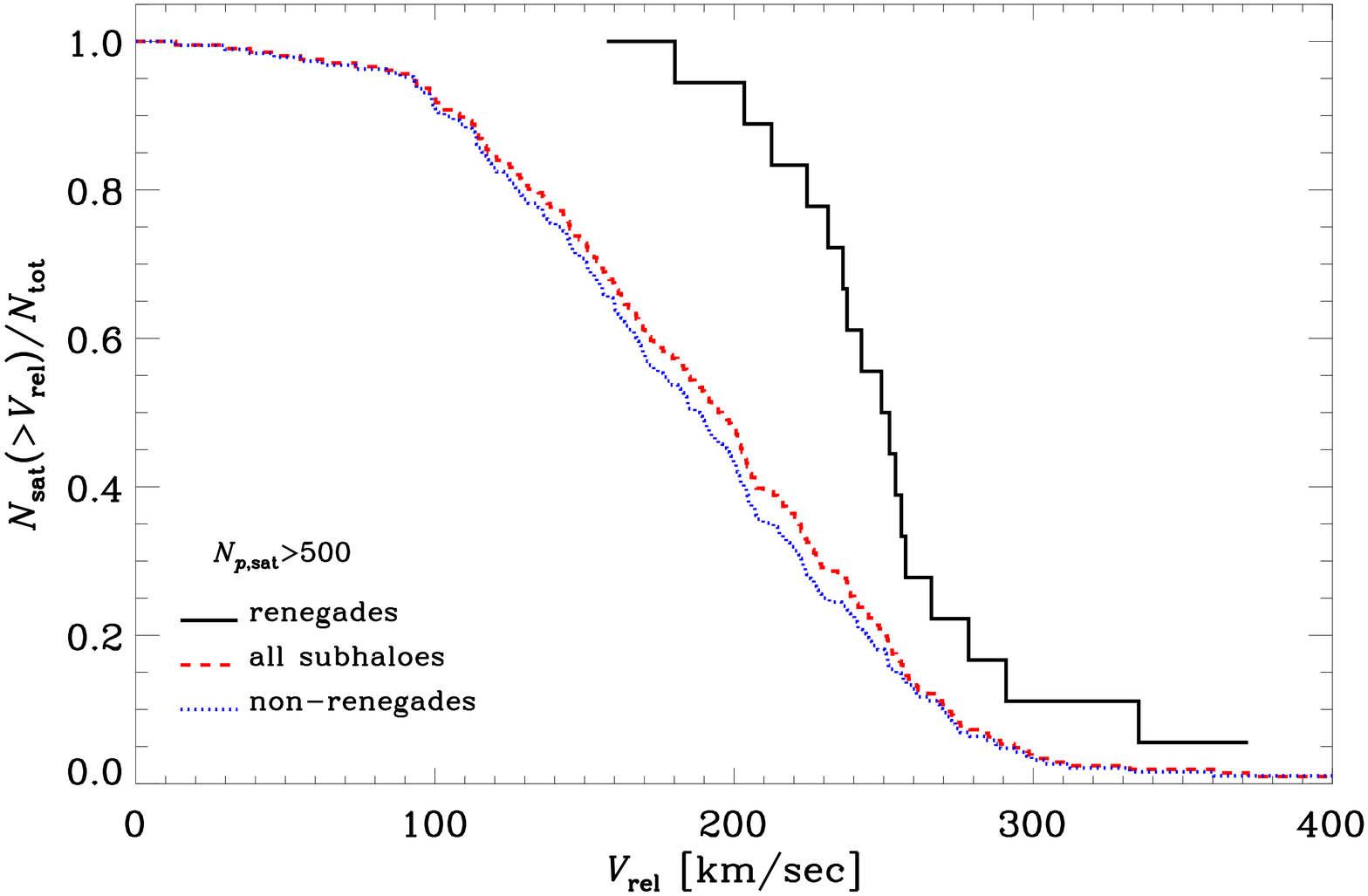,  width=0.40\hsize,height=0.24\hsize,angle=0}
\psfig{figure=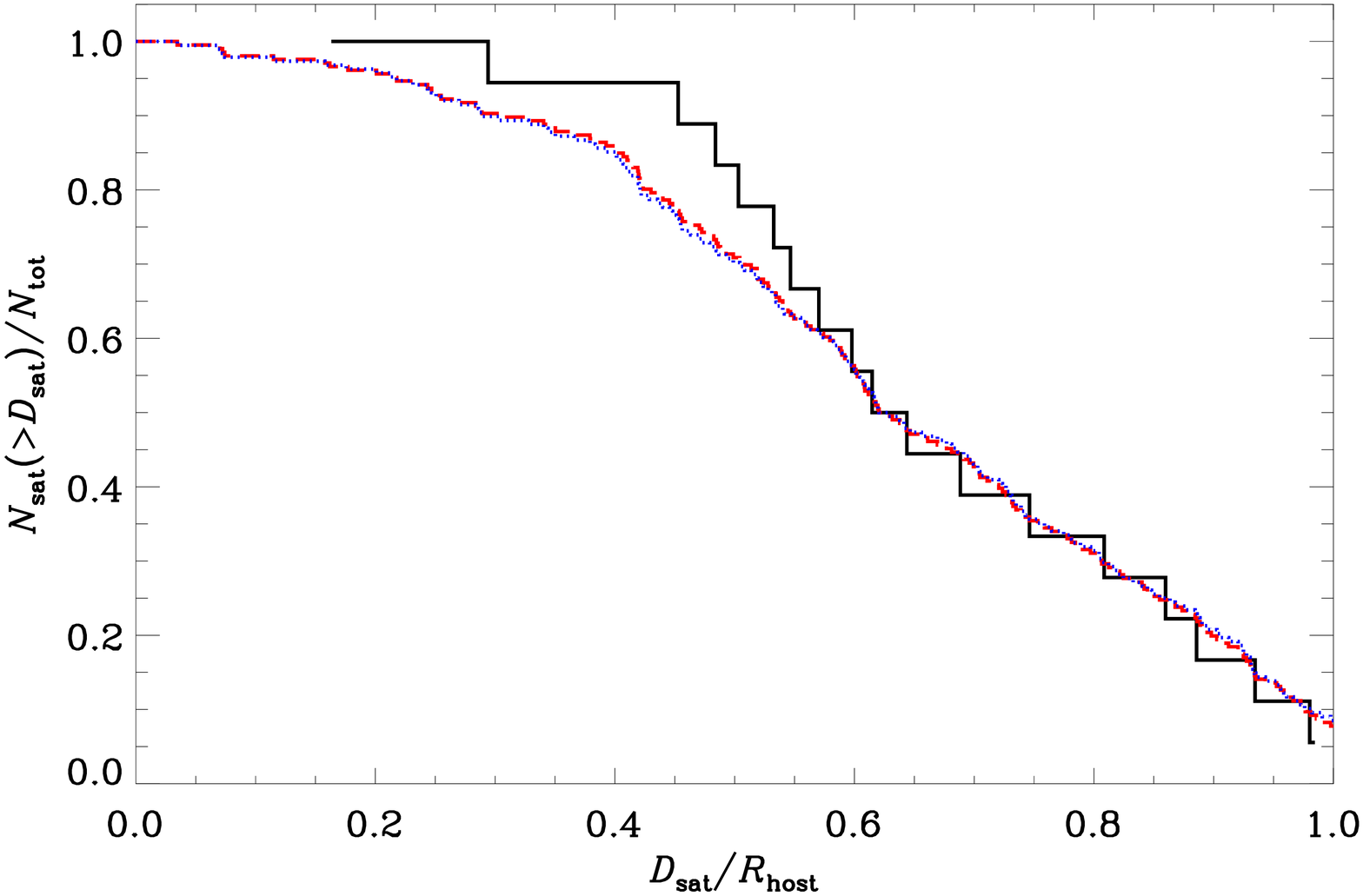,  width=0.40\hsize,height=0.24\hsize,angle=0}
}}
\centerline{\hbox{
\psfig{figure=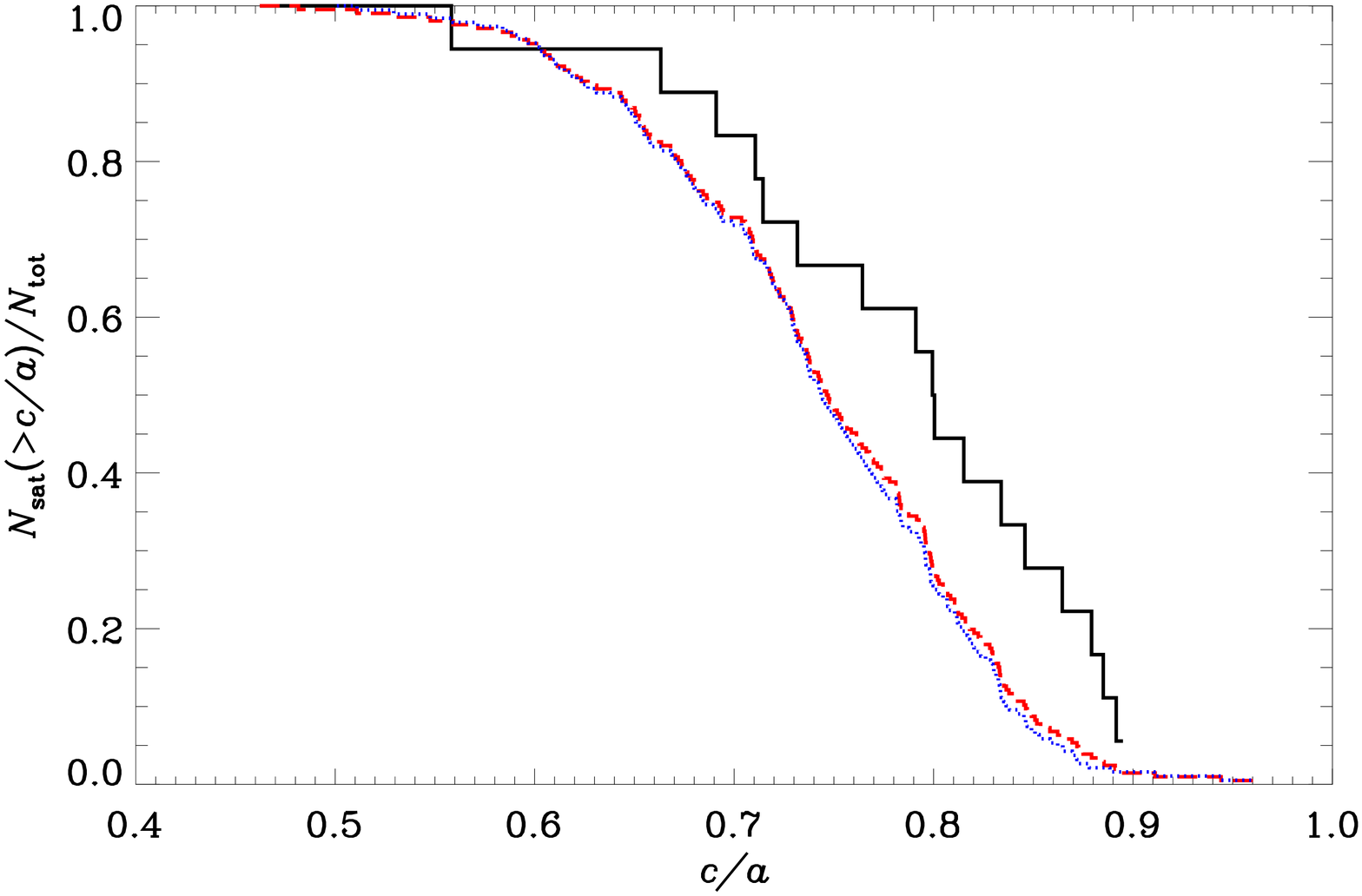, width=0.40\hsize,height=0.24\hsize,angle=0}
\psfig{figure=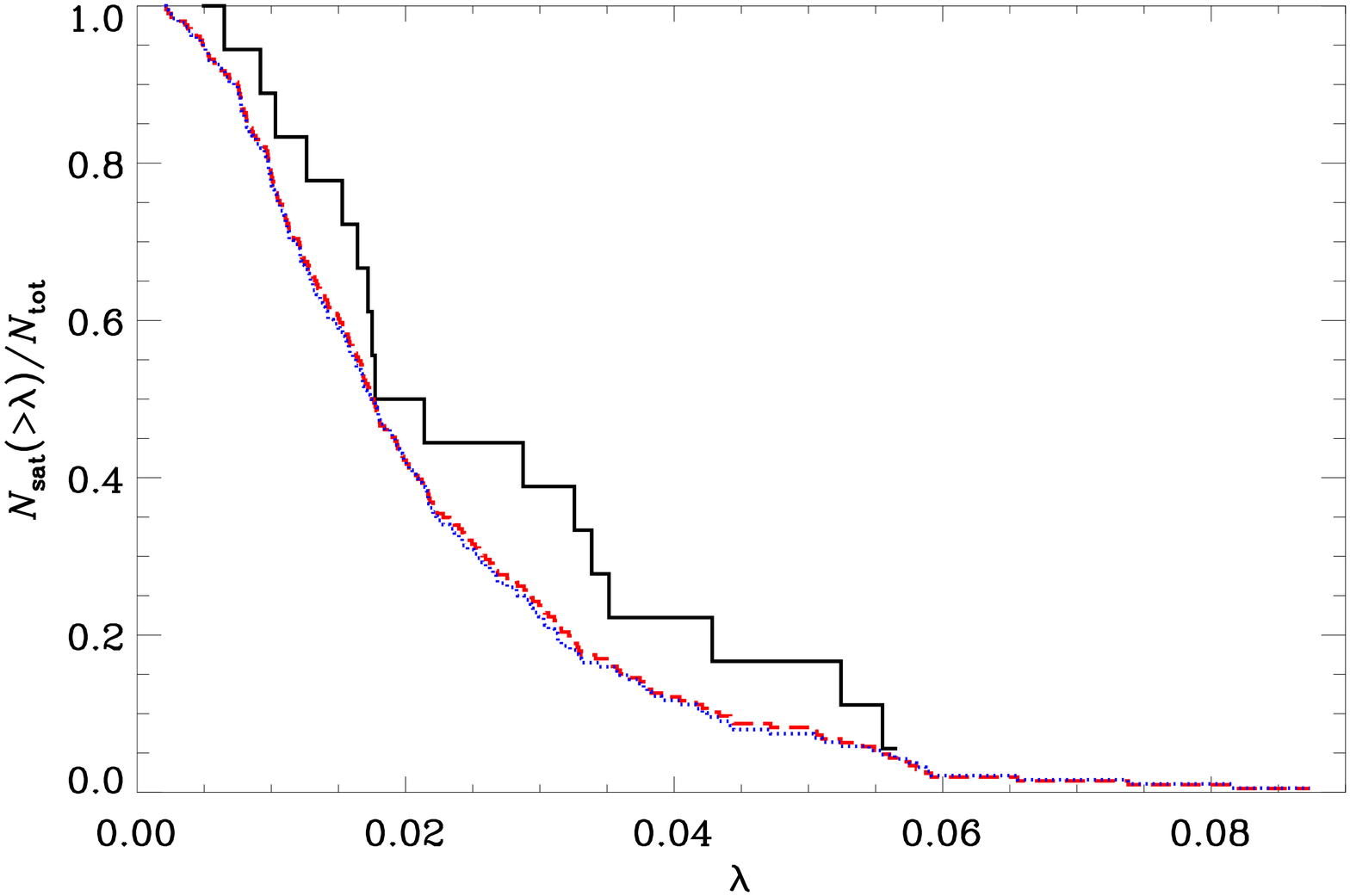,width=0.40\hsize,height=0.24\hsize,angle=0}
}}
\caption{Distribution of relative velocities (upper left), distance (upper right), sphericity (lower left), and spin parameter (lower right) for renegade subhaloes (solid) in comparison to all (dashed) and loyal (dotted) subhaloes at redshift $z=0$ combined for the MW and M31.}
\label{fig:PCompare}
\end{figure*}

We additionally query whether the unique past of renegade subhaloes will leave a noticeable imprint upon their (internal) properties other than their mass. To this extent we calculated several quantities of interest including their relative velocity (within the rest frame of their respective host), their distance to the final host at redshift $z=0$, their sphericity $s=c/a$ based upon the largest and smallest eigenvalue ($a$ and $c$, respectively) of the moment of inertia tensor \citep[cf.][]{Knebe10a,Knebe08}, their spin parameter $\lambda=|{\bf L}|/\sqrt{2}MVR$ \citep{Bullock01b}, and their concentration as defined by $x=r_{\rm sub}/r_2$ (where $r_{\rm sub}$ is the subhalo edge and $r_2$ the peak position of $\rho(r)r^2$ with $\rho(r)$ being the density profile). For this particular test and any possible comparison to the Magellanic Clouds we restricted all subhaloes to contain at least 500 particles leaving us with 18 renegades and 206 subhaloes in total. The solid lines in \Fig{fig:PCompare} refer to the distribution of the renegade subhaloes alone, while the dashed line represents the total subhalo population (including renegades); the dotted line represents just the non-renegade objects. Each curve is normalized by the respective total number of subhaloes.

From \Fig{fig:PCompare} we see that renegades posses substantially larger $z=0$ velocities. However, hardly any of the renegades show a velocity as large as the one observed for the Magellanic Clouds \citep[e.g. 340--380 km/sec, see][]{Kallivayalil06lmc, Piatek08}, likely because they do not enter as deep into the potential as their observational counterparts. Furthermore the most massive renegades are not the ones with the largest relative velocities. They also do not enter as deep into the potential as their loyal companions. However, we need to acknowledge that our simulation does produce non-zero objects with velocities comparable to the MC's even though their number is rather low. When examining the sphericity and spin of each renegade, we find that the they appear rounder and with a higher spin parameter than the general population. We examined the distributions of concentrations and found them to be indistinguishable from the full subhalo population (not shown here). Despite the apparent differences seen in \Fig{fig:PCompare} we nevertheless conjecture that they are unable to leave an discernible imprint on the respective distributions due to their low numbers.

\begin{figure}
\noindent
\centerline{\hbox{\psfig{figure=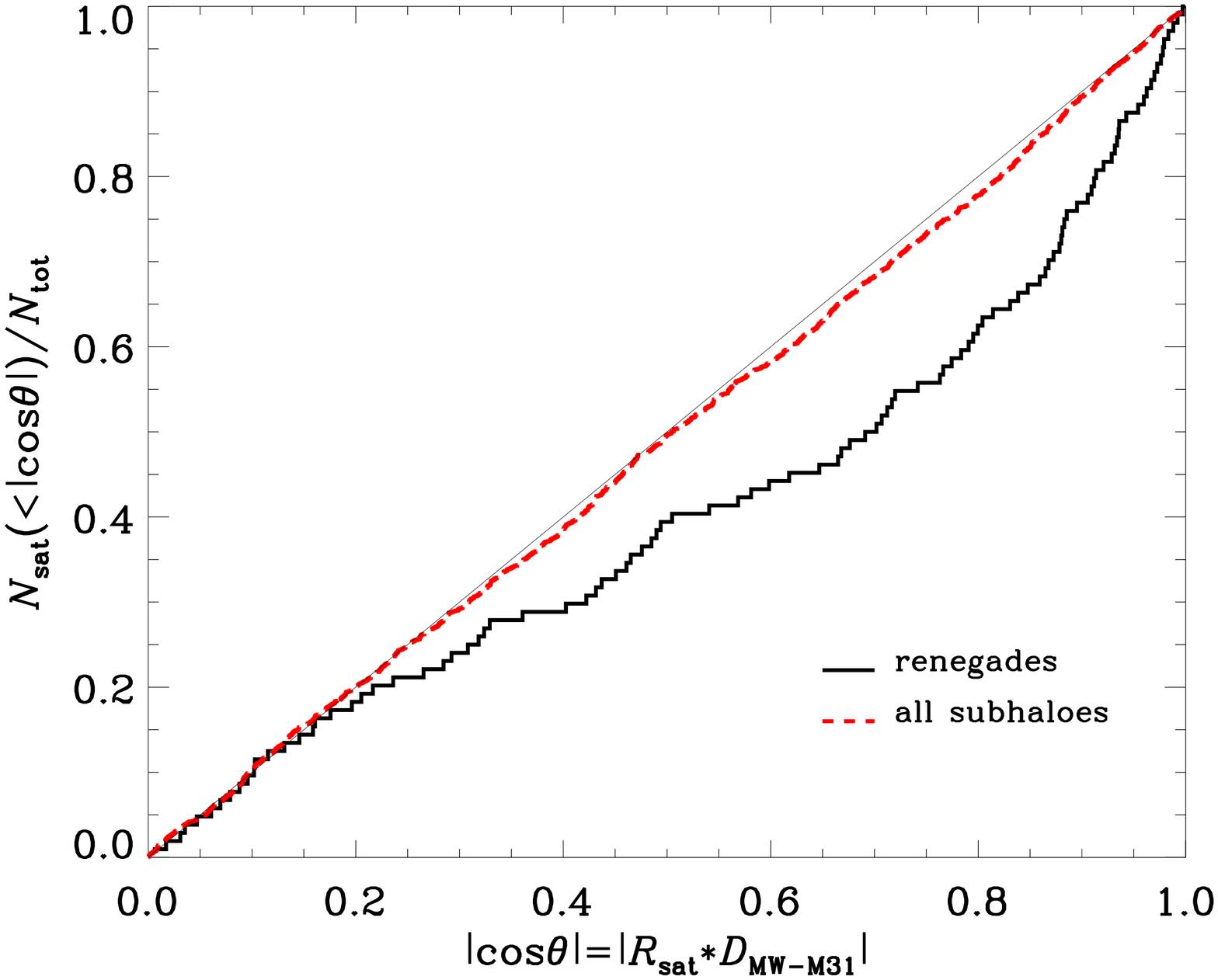,width=0.85\hsize,angle=0}}}
\caption{Distribution of the (cosine of the) angle between the position vector of a subhalo and the connecting line between the two hosts MW and M31.}
\label{fig:PthetaCompare}
\end{figure}

We finally examine the spatial distribution of our renegade subhalos by showing, in \Fig{fig:PthetaCompare}, the distribution of $\cos\theta={\bf R}_{\rm sat}\cdot {\bf D}_{\rm MW-M31}$ where ${\bf R}_{\rm sat}$ is the (normalized) position vector of a subhalo in the frame of its redshift $z=0$ host and ${\bf D}_{\rm MW-M31}$ is the (normalized) vector connecting the MW to M31. We can clearly see that this angle is isotropically distributed for the the full subhalo population of the two hosts, while, there is a clear preference for renegade satellites to be found in the direction of the other host. Please note that the near uniform distribution about ${\bf D}_{\rm MW-M31}$ found for the loyal satellites here does not contradict the generally accepted anisotropic distribution of subhaloes \citep[e.g.][]{Knebe04,Libeskind05,Zentner05,Libeskind07} as we have stacked the distributions from two hosts.

\section{Discussion and Conclusions}
\label{sec:concusions}
Motivated by recent claims that the two Magellanic Clouds \nil{were either formed close to} M31 \nil{(and ejected from it by} a recent merger) or simply accreted \nil{from the} outer reaches of the Local Group \citep{vandenBergh10, Busha10} we \nil{examined the likelihood of} this situation in Constrained Local UniversE simulations of the Local Group. We find that subhaloes disloyal to their first host exist, and called them renegade subhaloes. \nil{They are subhaloes accreted by one host halo at a given time, yet at $z=0$ are found within the virial radius of completely different host.} 

\nil{We have examined various physical properties of the renegade population and contrasted them with the full loyal population. We find that significant differences are seen in the relative velocities, sphericity, radial distribution and spin parameter. Furthermore, renegade subhaloes are significantly more anisotropically distributed than the full subhalo population - more renegades are found close to the line connecting our two main galaxies than not.}

Despite the differences \nil{highlighted above, detecting renegade subhaloes is only possible if they leave an imprint on the full subhalo population - its not enough to have vastly different properties.} One \nil{thus} has to compare the dashed \nil{with} the dotted line \nil{in the figures presented above, to gauge any observable signature. In principle,} an observer will only have access to the set of visible satellite galaxies and hence could generate a distribution akin to the ones presented in \Fig{fig:PCompare}. But only if loyal satellites show a substantially different behavior to the combined sample will there be a chance to (observationally) confirm their existence \citep[as, for instance, the difference in the velocity distribution for backsplash and infalling subhaloes reported by][]{Gill05}. We thus conclude that their (observational) detection will be difficult unless orbital information can act as a discriminator.

%

The \nil{abundance} of renegade subhaloes \nil{appears to be} a function of \nil{both} the distance between the hosts "sharing" them (which also corresponds to redshift in our case) and the mass of the hosts. \nil{Yet despite their existence, it is still a great challenge to explain the origin of the MCs as renegade subhaloes. In our (dark matter only) simulation, renegade subhaloes are not massive enough to realistically be called ``MC''-like. Our inability to find a MC-like renegades, however, may simply be due to the difficulty in producing such massive satellites in the first place. Further, our simulation currently only considers dark matter whereas it has been shown before that baryonic physics will have a certain impact upon the properties of subhaloes \citep[e.g.][]{Libeskind10,Romano-Diaz08}. Future work that quantifies the frequency of finding MC-like objects in MW sized haloes, and extends this work by then examining the likelihood that these subhaloes become renegades is needed to verify or falsify this formation scenario.}

\section*{Acknowledgements}
AK is supported by the MICINN in Spain through the Ramon y Cajal programme as well as the grants AYA 2009-13875-C03-02, AYA2009-12792-C03-03, and CAM S2009/ESP-1496. We thank DEISA for granting us supercomputing time on MareNostrum at BSC and in SGI- Altix 4700 at LRZ, to run these simulations under the DECI-SIMU-LU and SIMUGAL-LU projects. We also acknowledge the MultiDark Consolider project CSD2009-00064 and the ASTROSIM network of the European Science Foundation for the financial support of the workshop "CLUES workshop" held in Brighton in June 2011 where this paper has been finished. GY acknowledges financial support from FPA 2009-08958, AYA 2009-13875-C03-02 and CAM S2009/ESP-1496, too. YH has been partially supported by the ISF (13/08).

\bibliography{archive} \bsp

\label{lastpage}

\end{document}